# Robust and customized methods for real-time hand gesture recognition under object-occlusion

*Abstract*—Dynamic hand tracking and gesture recognition is a hard task since there are many joints on the fingers and each joint owns many degrees of freedom. Besides, object occlusion is also a thorny issue in finger tracking and posture recognition. Therefore, we propose a robust and customized system for realtime hand tracking and gesture recognition under occlusion environment. First, we model the angles between hand keypoints and encode their relative coordinate vectors, then we introduce GAN to generate raw discrete sequence dataset. Secondly we propose a time series forecasting method in the prediction of defined hand keypoint location. Finally, we define a sliding window matching method to complete gesture recognition. We analyze 11 kinds of typical gestures and show how to perform gesture recognition with the proposed method. Our work can reach state of the art results and contribute to build a framework to implement customized gesture recognition task.

## I. INTRODUCTION

REAL-TIME hand tracking and recognition has become an attractive topic based on that ubiquitous vision based sensors are well spread all over the world, and it's especially important in interaction-based industrial applications, such as Microsoft Hololens, Apple ARKit , other applications in simulation robot interaction etc. Efficient hand interaction can ensure the natural communication between human and computers. Every camera can perform as a basic input sensor in machine perception. There have been extensive studies on vision based hand gesture recognition, such as [1]–[4]. These methods have applied in various environments with effective and promising results. However, many of them focused on kinematic pose estimation of an isolated hand. The gesture recognition with occlusion and finger self-occlusion have not been covered throughly [5]. The challenge of dynamic hand recognition is the identification of hand keypoints sequence. The most common assumption for hand identification is that a single hand owns 16-20 keypoints, 25-50 DOF(Degree of Freedom) in total [6]–[8]. Each keypoint needs to quantify the relationship between the current and its former state(e.g., coordinate,angle) in time series. Real-time gesture recognition in occlusion is even more difficult generally for three reasons. (i) Some keypoints attributes cannot be captured. This is a big problem for most of the model-based methods. (ii) Missing data. This will lead to discontinuous time sequences, and further complication to maintain consistency of the keypoints context states. (iii) Insufficient observed sequences generate severe time malposition to the defined gesture templates. This can make the hand states be classified into wrong category.

Occlusion creates obstacles for finger detection and recognition. The possible solution is to supplement missing articulations or other keypoints in real-time 3D hand gesture assessment process. Many trials have done to predict missing data. Recently, some state-of-art research [5], [9] adopted joint-aware principle, considering joint pattern and mutual correlation. It used the local keypoints mode to formulate different forecasting and discriminating strategies. At the same time, Neural Networks [10]–[12] are widely used in most of their works, especially Convolutional Neural Network (CNN) [13], [14]. CNN can be used to effectively extract different features in each gesture, and quantify the relationship between the most relevant features and estimated hand postures. Besides, there are some improved algorithms based on CNN, such as 3DCNN, as introduced in [15], [16]. They also got excellent results. In addition, model-based methods are also proved to be very effective. Zhou Xingyi et al. [17] proposed a deep learning based approach. It improved greatly in hand tracking and gesture recognition accuracy. These methods can reduce the impact of occlusions in specific conditions.

Although deep learning based methods have achieved encouraging results, their experiments are running in limited training datasets. A perfect training dataset is difficult to find. Training sets is as important as model architecture, this is discussed in [18]. Many studies utilized open source hand datasets, such as *NY UHandPoseDataset*, *HandNet* and *Dexter*1, or collected gesture data through webCam, Kinect or other data acquisition equipments. Using already existed datasets limits in pertinence in practice, it's difficult to mark the labels specified for customized algorithms one by one. But the collection of a large group of hand gestures is a complicated process. In this study, we present a new structure based on Generative Adversarial Network (GAN) to collect a small portion of hand gesture data and generate more approximately real ones. This method can effectively speed up the data production and reduce time cost.

The contributions of our paper are as follows.

1) We propose a RNN-NN connected dataset generation model based on the principle of GAN. This model can generate approximate real discrete data efficiently.
2) We model the angles between hand keypoints and encode their relative coordinate vectors according to kinematics, combining the probability distribution to complete gesture model.

3) We propose a time series forecasting method to predict missing data and define a sliding window matching method to capture correct keypoint sequences from realtime input stream.

The paper is organized as follows. The related works includes current occlusion solutions and relevant methods used in this work are introduced in Section II. The definition of the angles between hand keypoints and their relative coordinate vectors are given in Section III. The framework and proposed methods are presented in Section IV. Furthermore, in Section V and VI we give the experimental setup and results. Finally, we conclude the paper in Section VII.

## II. RELATED WORK

Data generation Some researchers have noticed the data problem. Markus Oberweger et al. [19] proposed a semiautomated method for efficiently and accurately labeling each frame of a hand depth video with the corresponding 3D articulation locations, it can spur the creation of accurate 3D hand poses annotations. Simon Tomas et al. [2] presented a multiview bootstrapping method, which used reprojected triangulations reflected from noisy detections as training data. However, typical GAN based methods used to generate discrete sequence data have not been well studied and applied. Occlusion solutions Various methods are conducted to reconstruct missing data or avoid data loss essentially [5], [10]–[12], [14]–[16]. Chiho Choi et al. [5] trained a pair of object oriented network and hand-oriented network using paired depth images to create a more informed representation. L. Ge et al. [13] project 3D hand points onto three orthogonal planes to generate heat-maps for three views. Multi-perspective fusion based methods can also be found in [20], [21]. These methods can help to maintain some hand joints data. However, the data is always insufficient. In those inevitable occlusion cases, many recent works adopted model-based approaches. I. Oikonomidis et al. [22] formulated an optimization problem, in which they combined the 26-DOF hand pose and model parameters of the manipulated object. Model-based principle contains the basis of template matching procedure in our work. Sequence prediction Sequence prediction is frequently used to make up for missing data or predict specific gesture. Typical methods include RNN, LSTM, as introduced in [23], [24]. Recently proposed recurrent convolutional network (RCN) has been adopted in [25]. It described a class of recurrent convolutional architectures which is end-to-end trainable and suitable for large-scale visual understanding tasks. Pavlo Molchanov et al. [26] employed connectionist temporal classification to train the network to predict class labels from in-progress gestures in unsegmented input streams. Further work combined 3DCNN with Convolutional LSTM that stated in [27]. Our proposed method is similar with [28]. It translates the 3D hand and human pose prediction problem from a single depth map into a voxel-to-voxel prediction which uses a 3D voxelized grid and estimates the per-voxel likelihood for each keypoint. We use single depth images and perform keypoints prediction, the difference lies in that we model the keypoint coordinate vector and interaction angles, and predict the missing data based on previous states.

## III. MODEL REPRESENTATION

In this section, we first decompose an isolated hand. Then the interaction angle model and keypoint coordinate vector model are established based on defined coordinate system.

### A. Hand decomposition

As illustrated in Fig.1(a), we assume there are 20 keypoints in a static hand, categorized into 6 classes in total. This can be described as $\Theta = \{A,B,C,D,E,F\}$. $ClassB,C,D,E$ each owns 4 keypoints, numbered as $0,1,2,3$ from top to down respectively. $ClassA$ owns 3 keypoints, numbered as $0,1,2$. $ClassF$ has only one element, denoting hand elbow.

To determine the relative position of hand keypoints, we adopt three keypoints $\{\Psi_0,\Psi_1,\Psi_2\}$ to form a 3D orthogonal coordinate system. Here $\{C_3,B_3\}$ are used as $\{\Psi_1,\Psi_2\}$. $\{F_0\}$ is used as $\{\Psi_0\}$. These three keypoints are always on a plane $\Phi$ for most hand gestures. The line $L_Y$ in which $\Psi_0$ and $\Psi_1$ are located denotes axis $Y$ and the line $L_X$ perpendicular to $L_Y$ within $\Phi$ denotes axis $X$, the line $L_Z$ through $\Psi_0$ perpendicular to $\Phi$ denotes axis $Z$.

### B. keypoint coordinate vector model

As shown in Fig.1(b), we take $ClassE$ and $ClassF$ for example. For a static state $S_t$, there are 4 coordinate vectors, denoting $\vec{V}_t(E0), \vec{V}_t(E1), \vec{V}_t(E2), \vec{V}_t(E3)$. The change of one vector represents the performance of a particular dynamic gesture on this feature, shown in Eq.(1).

$$\Delta \vec{V}_t(E0) = \vec{V}_t(E0) - \vec{V}_{t-1}(E0) \qquad (1)$$

Also, $\Delta$ can be decomposed into features in these three directions, denoted in Fig.1(d). For those missing keypoints, the previous states are used to predict the current status based on proposed forecasting methods.

### C. Angle model

Based on predefined keypoint coordinate vectors, interaction angles depend on three nearest neighbors. The static state $\vec{S}_t$, $\alpha_t(E012)$ can be expressed by two vectors $\vec{V}_t(E2E1), \vec{V}_t(E1E0)$, as in Fig.1(c).

## IV. FRAMEWORK AND METHODS

In this section, first we present the overall framework of the proposed system, then introduce individual method used in it. Meanwhile, data processing and flow will be described.

### A. Overall Framework

As shown in Fig.2, the overall framework contains three modules, data generation, sequence learning and gesture recognition. Data generation adopts a RNN-NN network to generate discrete sequences. Let $N_{sample}$ represent the number of samples for each gesture. Generated data are stored as original datasets. Sequence learning uses Gated Recurrent Unit (GRU) to perform feature extraction and status prediction. Obtained sequence features are saved as gesture model after several steps of optimized processing. Gesture recognition and judge the current gesture state using joint probability distribution.

*B. Vector Encoder*

Due to the spatial diversity of human hands(e.g. distance to the camera, hand size), pixel distance cannot be used directly shows how to recognize hand gestures from real-time depth to define the training data. Therefore we put forward a input stream with occlusions.

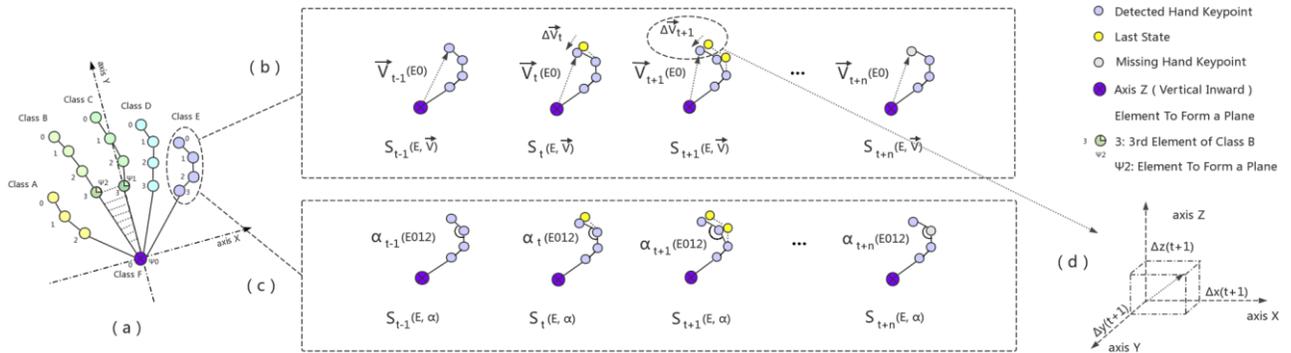

Fig. 1. Isolated hand decomposition. (a) 6 kinds of classifications of hand keypoints, with 3D spacial orthogonal coordinate system definition. (b) Intersection angles. (c) Keypoint coordinate vector. (d) Displacement decomposition.

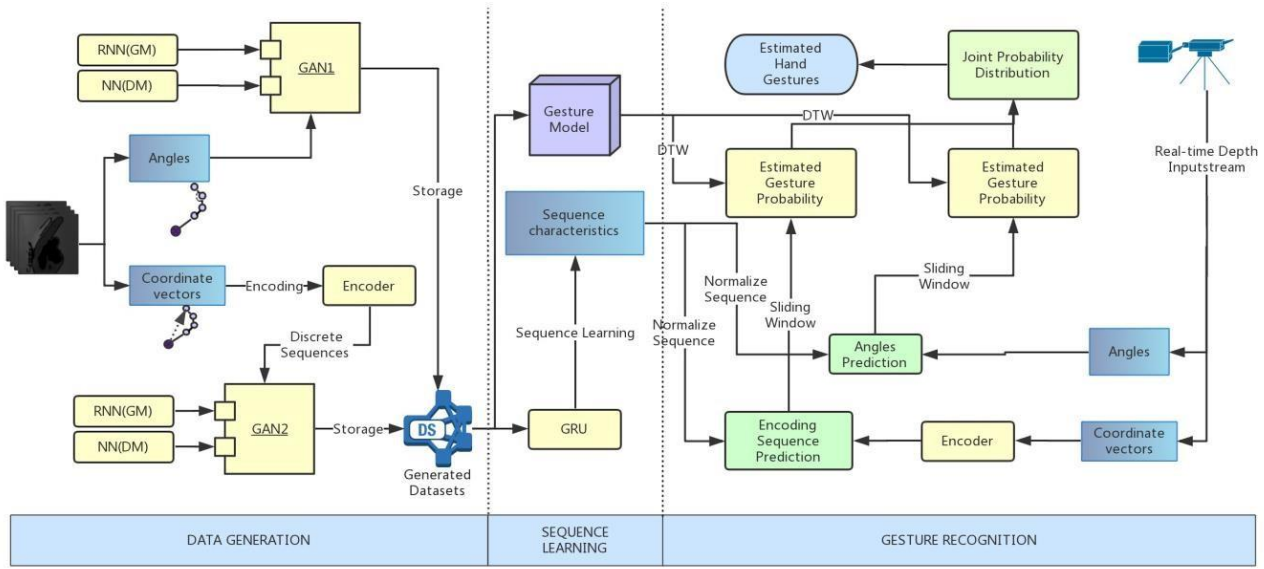

Fig. 2. Overall framework of the proposed system, including (1) Data generation, (2) Sequence learning, (3) Gesture recognition.

First, non-holonomic coordinate vectors and angles due to occlusions are extracted, undetected angle sequences are directly predicted using learned sequence characteristics, while the coordinate vectors are encoded into three orthogonal components to perform prediction. Next a sliding window is called to intercept corresponding gesture segments. We fit the segmented data to the derived model in order to obtain the estimated gesture probability distribution respectively. Finally, we fuse these two types of probabilities, vector encoder to transform pixel distances into scale invariant discrete scalars. The encoder has two functions.

Fig. 3. (a)-(g) Keypoint extraction and vector encoding. (h)-(i) Distance and confidence distribution heatmap.

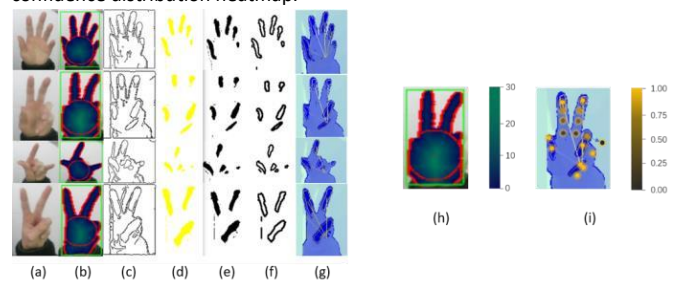

- Find the gravity center of the palm $H_{center}$, we assume the length from the elbow $F_0$ to $H_{center}$ as the baseline length $\Gamma_{baseline}$. $H_{center}$ can be derived using Eq.(2).

$$H_{center} = \operatorname*{argmax}_{p}(\operatorname*{argmin}_{p,q}|p - q|_{Euc}) \quad (2)$$

We use a RGB skin detector to extract hand area then perform binaryzation, $p$ denotes point on the ROI(region of interest) and $q$ denotes point on the background. Set the minimum Euclidean distance from $p$ to $q$ as a temporary value of $p$, while the $p$ owning the maximum value represents $H_{center}$.

- Normalize all of the pixel distances using the ratio to derived $\Gamma_{baseline}$. Therefore, vectors can be encoded into three discrete values and the consistency is maintained.

Detailed process is shown in Fig.3. we perform on 4 kinds of typical hand gestures. Here (a) denotes original hand color map, while (b), (c)-(g) are conducted under RGBA map and depth map respectively. The red point in (b) expresses the gravity center $H_{center}$ and the red circle expresses the size of palm. The minimum distance of $p$ to $q$ corresponding to a certain RGBA value is defined, as shown in Fig.3(a). The distance distribution of $p$ could be observed intuitively in gradient color diagram (b), which is vital in vector composition. (c) shows extracted hand contour from depth map. We set a threshold and segment several prospect regions $R_{pros}$ in (d). As for points on $R_{pros}$, their encoded distances to the background depend on the size of $R_{pros}$. To obtain hand keypoints position, we perform binaryzation on (d) and get connected domain.

Take the center of each connected region $R_{pros}$ with relative position of $H_{center}$ as the coordinate vector of region $R_{pros}$, shown in Fig. 3(g). However, not all pre-defined keypoints could be extracted. For those prospect keypoints, we have a high confidence on their positions, while a relatively lower confidence on the other keypoints which cannot be segmented directly. Here we introduce a simple hand skeletal framework to roughly quantify the confidence (spans [0,1]) of the keypoint coordinates. This confidence represents the proportion of this keypoint in sequence prediction. The fitting process is very simple, assign the coordinates of the points in the extracted prospect regions to corresponding points in the skeleton. Typical distribution is illustrated in Fig.3(i).

As shown in Fig.3(i), the points on the prospect regions own 1.0 confidence. The value decreases with the distance from the foreground, because the uncertainty of a keypoint's position is positive correlated to the distance from the prospect regions. Derived baseline $\Gamma_{baseline}$ maintains good scalability, since all Euclidean distances wouldn't exceed 4 times of $\Gamma_{baseline}$. Therefore confidence $Conf$ can be calculated using Eq.(3).

$$Conf = 1 - D_{Euc}/(4 * \Gamma_{baseline}) \quad (3)$$

Where $D_{Euc}$ denotes the minimum Euclidean distance of a keypoint to prospect region in the same class. Since those occluded keypoints' coordinate could be approximately estimated using skeleton framework, we set their $Conf$ initially as 0. It has to be adjusted according to the sequence prediction.

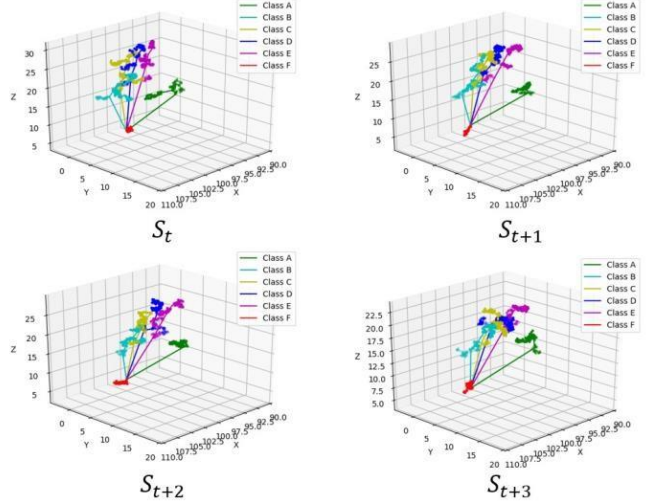

Fig. 4. 4 adjacent keypoint state sequences.

### C. Generate data

Our proposed GAN module employs RNN as the generative model and a one-layer neural network as the discriminative model. RNN is promising in generating contextsensitive sequences, and NN is also effective in classification. Kinect provides real-time RGBD data of $M$ frames per second. We collect $N$ segment data per gesture as initial samples. Therefore, each gesture can be expressed using $C_N\{S_M\{\Theta,\Delta\alpha\}\}$ or $C_N\{S_M\{\Theta,\Delta V_x,\Delta V_y,\Delta V_z\}\}$

$$\rightarrow -$$

encoded from $C_N\{S_M\{\Theta,\Delta V\}\}$. $C$ denotes collection state. Initial samples are used as ground truth data. As for generative network, we employ two random noises $Noi_\alpha$, $Noi_v$ for angle based model and vector based model respectively according to different state step length. Input noise variables are shown as Eq.(4).

$$\lambda = \begin{cases} C_N\{S_M\{\Theta,\Delta\alpha\}\} + rand(\pm 5) & Noi_\alpha \\ C_N\{S_M\{\Theta,\Delta V\}\} + rand(\pm 1) & Noi_v \end{cases} \quad (4)$$

Where $rand(\pm a)$ denotes a uniform distributed random real number between $-a$ and $a$. We set 5 for $Noi_\alpha$ and 1 for $Noi_v$ in this paper. But they can be customized according to different conditions. Based on general GAN framework, generator can be optimized by minimizing Eq.(5) and discriminator can be optimized by maximizing Eq.(6).

$$E_{\lambda \sim p_\lambda(\lambda)}[log(1 - D(G(\lambda)))] \quad (5)$$

$E_{x \sim p_{data}(x)}[log(D(x))] + E_{\lambda \sim p_\lambda(\lambda)}[log(1 - D(G(\lambda)))]$ (6) $p(\lambda)$ is a prior distribution of $\lambda$, generator $G(\lambda)$ maps from the space of $\lambda$ to the space of data $x$. $D(\cdot)$ is the discriminator with a [0,1] output. The generated discrete data are stored as training sets.

Fig.4 shows 4 adjacent keypoint state sequences, each contains 40 groups of coordinate data selected from the generated set. As we can see, the position distribution of the fingertips and the elbow is relatively concentrated, while the other keypoints distribute in scattering. There are two main reasons for this result. First, the fingertips and the elbow own high position confidence, while the others not. Second, GAN generates data close to real distribution, coordinate error will not be reduced.

### D. Sequence Learning

As for raw datasets, a gesture contains $M \times 14$ (14 $\Delta \alpha$) and $M \times 19 \times 3$ (19 $\Delta v$ and 3 components). Each variable has its own state shift mode in $M$ frames. Nevertheless, variables of other classes are also related to the estimated state one, especially those in the same class. In order to quantify this relationship, we define some assumptions. First, the variables not in the same class(e.g.$\Delta E0$ and $\Delta F0$) has no effect on prediction. Second, the variables that owns the same class and same state have an IF (influence factor) of 0.5. Third, the variables in the same class but different states have an sum average IF as 0.5. Let $\beta$ denote IF, $\beta_0$ denote IF of the main stem, then we train the network using Eq.(7)-Eq.(11).

$$z_t = \sigma(W_z \cdot [h_{t-1}, x_t(\Theta_{j,i})]) \quad (7)$$
$$r_t = \sigma(W_r \cdot [h_{t-1}, x_t(\Theta_{j,i})]) \quad (8)$$
$$\tilde{h}_t = tanh(W_{\tilde{h}} \cdot [r_t * h_{t-1}, x_t(\Theta_{j,i})]) \quad (9)$$
$$h_t = (1 - z_t) * h_{t-1} + z_t * \tilde{h}_t \quad (10)$$
$$y_t(\Theta_{j,0}) = \sigma(W_0 * \beta_i * y_t(\Theta_{j,i})) \quad (11)$$

Eq.(7)-Eq.(10) are general GRU network. $z_t, r_t, \tilde{h}_t, h_t$ denote update gate, reset gate, candidate activation and activation layer respectively. $\Theta_{j,i}$ denotes the $i_{th}$ keypoint in the $j_{th}$ class of $\Theta$, and $\beta_0 = 0.5$ when $i = 0$.

Based on the trained sequence prediction model, a missing keypoint at a certain state can be estimated by several of its previous states, and some states of other keypoints in the same class, as illustrated in Eq.(12).

$$S_{missing} = \sum_{len(\Theta_j)} \beta_i \cdot Model(S_i) \quad (12)$$

Here $\beta_i$ denotes weighted influence factor, and the state $S$ includes discrete coordinates and angles.

### E. Normalize gesture sequence

We assume that a complete gesture contains 30 frames. However, it's very difficult to finish a gesture exactly using only one second. Indefinite-length actual gesture sequence cannot perform gesture recognition because each frame is not a oneto-one correspondence. Here we propose a method based on the longest common substring to normalize actual gesture sequences. First we encode coordinator vector and angle into specific categories, we adopt 0.2 for coordinator vector and 20 degree as step length. For example, if a keypoint owns $(0.18, 1.06, 0.92)$ and $(144°)$, it will be encoded as $(A_v, F_v, E_v)$ and $(H_a)$. The whole encoding relationship is shown in Table I. The first and forth line denote the range of coordinator vector. The second and fifth line denote the range of angles, the third and sixth line are encoded categories. All of the experiment data are within the maximum value we settle, which can ensure no invalid encoding definition.

TABLE I ENCODING RELATIONSHIP

| 0~0.2 | 0.2~0.4 | 0.4~0.6 | 0.6~0.8 | 0.8~1.0 |
|---|---|---|---|---|
| 0~20° | 20~40° | 40~60° | 60~80° | 80~100° |
| A | B | C | D | E |
| 1.0~1.2 | 1.2~1.4 | 1.4~1.6 | 1.6~1.8 | 1.8~2.0 |
| 100~120° | 120~140° | 140~160° | 160~180° | 180~200° |
| F | G | H | I | J |

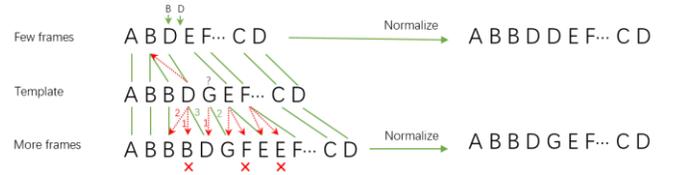

Fig. 5. Longest common substring normalization.

There are mainly three types of cases, actual gesture contains fewer frames, same frames and more frames than template. As for the same frame cases, we can use the frames data directly. As for the other two cases, we normalize the sequence using proposed method shown in Fig.5. When a gesture is performed slowly and contains more frames, captured realtime sequence shows like "More frames". First "A","B","B" in the template can find corresponding unit in the "More frames". But "D" cannot hit a corresponding "D" in the same position. Then we try to hit "D" in the last recorded index of "More frames", shift to the next position until the hit is successful and we record this hit index. As for "G", it can't hit success in the same position either. The current recorded index is "D", so shift to the next position until "G" can hit success. When a gesture contains fewer frames, typical example shows like "Few frames". As for the second "B" in the template, it cannot hit success, so hit the last recorded index and succeed.

In this case, insert a new "B" in the corresponding position. As for "G", it does not exist in the captured sequence, insert a new element in the same position and the element is the last recorded element. Traverse the template according to this rule and we can get output like "ABBDGEF...CD" and "ABBDDEF...CD", it is different with these two outputs. Even though, normalized sequences are much more similar to the templates compared to captured real-time sequence.

*F. Dynamic matching and gesture recognition*

Matrix sequence could be completed using trained prediction model. However, current state and its last $M-1$ states cannot be used to perform gesture recognition since they may belong to two different gestures. Here we propose a sliding window method to match candidate sequence with the defined gesture templates dynamically. First, we catch $1.5M$ frames of data and set that more than $0.5M$ frames are required to identity a gesture. Then we use sliding window to derive joint probability distribution of mean error, recognize real-time gesture. Detailed recursive matching process is given in Algorithm.1.

---

**Algorithm 1** Recursive sliding window algorithm
**Require:** $1.5M$ frames of data
**Ensure:** Identified hand gesture

1: Initialize cursor=-M, error $E$, mean error $ME$, map $V$
2: repeat
3: $$E = \sum_{i=-1.5M}^{cursor} \sum_{j=0}^{dim} Conf_{i,j} * |Candi_{i,j} - Tmpl_{i,j}|$$
4: $ME = \frac{1}{K*N_{dim}} * E$
5: $K = \begin{cases} 1.5M + cursor & -M \leq cursor \leq -0.5M \\ M & -0.5M \leq cursor \leq 0 \end{cases}$
6: Searching templates for corresponding gesture of the minimum value $ME$, push them into $V$, let cursor++
7: until cursor=0
8: Define relative error $RE$, iterate through $V$
9: $RE_i = \frac{ME_i}{\sum_0^{size(V)} ME_i}$, calculate the relative error under two strategies(coordinate and angle), the gesture owning the minimum value is identified as current gesture

---

Here $N_{dim}$ denotes the dimension of a frame, and values 19x3 for coordinate vector and 14 for angle respectively. $Conf_{i,j}$ denotes the confidence of $j_{th}$ sub-state in the $i_{th}$ frame as mentioned in Sec.IV-B. *cursor* denotes a supplementary variable, varying from $-M$ to $0$, ensuring more than $0.5M$ frames of data are used to recognize hand gestures.

The process stores the past $1.5M$ frames of raw data relative

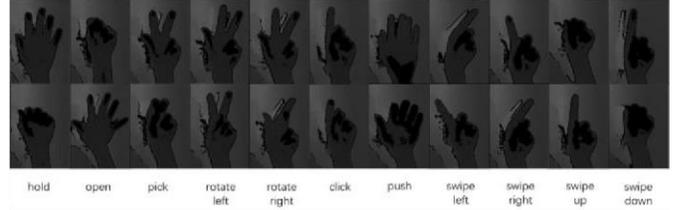
hold  open  pick  rotate left  rotate right  click  push  swipe left  swipe right  swipe up  swipe down to current state, then traverse *cursor* from $-M$ to $0$. This is the kernel of this sliding method. As for each state, we use the minimum average Euclidean distance between $K$ raw data and last $K$ states of gesture templates to represent mean error, and save the minimum mean error and the corresponding gestures. Two groups of error data can be obtained at the end of the traversal. In order to maintain the judgment consistency under coordinate and angle group, we calculate the proportion of each absolute mean error in the total error. Find the minimum relative error, its corresponding gesture denotes the current destination gesture.

V. EXPERIMENT

In this section, we give an example to apply our method, and introduce the detailed experiment setup.

*A. Experimental Design*

As we mention in the Sec.IV, there are some undetermined parameters, here we set M=30, $N_{dim}$=19x3 for coordinate and =14 for angle, $N_{sample}$ = 50.

*B. Experiment setup*

As illustrated in Fig.6, we define 11 types of representative hand gestures, which covers typical finger characteristics. Fig. 6. Defined 11 kinds of representative hand gestures..

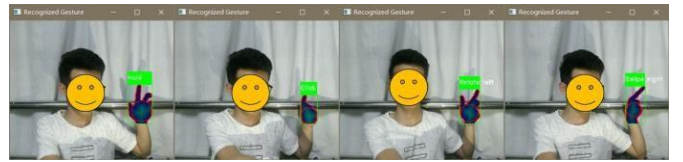
Fig. 7. Real-time gesture recognition results.(a)hold,(b)swipe right,(c)rotate left,(d)click.

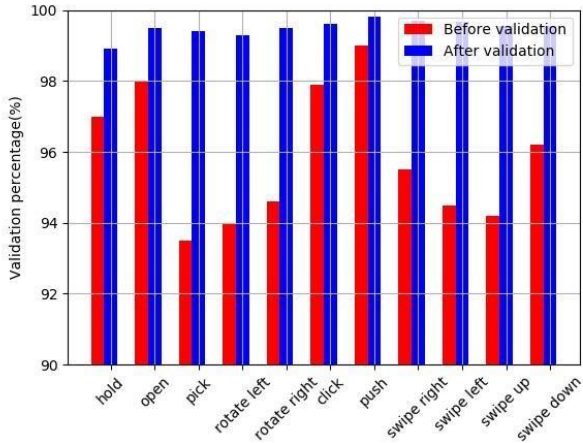

Fig. 8. Dataset validation of defined gestures.

First, we collect 50 samples for each gesture, extract coordinate vectors and angles as raw dataset, then use two defined GANs to generate $5 \times 10^3$ samples respectively as training set. The preprocessing phase includes keypoint information extraction and discrete data generation, while the sequence learning part includes discrete state sequence learning and model optimization. The model training process to coordinate vector and angle are in parallel. We use a GPU(NVIDIA GTX1050) to accelerate real-time RGB and depth data calculation, to ensure 30FPS.

## VI. RESULT

In this section, we present experiment results, including recognition accuracy under different conditions, comparison with other recent methods, to show superiority of our proposed method.

### A. Real-time gesture recognition result

As shown in Fig.7, we give 4 typical gesture recognition results, (a)-(d) represent hold, swipe right, rotate left and click corresponding category. Update the center of each categories and repeat until all sequences are traversed. Detailed algorithm is shown in Algorithm.2. Compared to the red histogram in Fig.8, we can see that dataset is much more accurate after validation.

approach.

respectively.

---

2:     Initialize 11 center clusters $C_i$
3: Take one record sequence $Seq_{cand.}$ and calculate distances to $C_i$, update $Seq_{cand.}$ to corresponding category of the minimum distance
4:     Update 11 center clusters $C_i$
5: until All record sequences are traversed

---

### C. Self-Comparison

*1) Non-Normalization vs Normalization:* First we try with non-normalization approach. We assume that a hand gesture is composed of 30 frames and implement our method strictly based on 30 frames of data. However, actual hand gestures

**Algorithm 2** Validation algorithm
**Require:** Raw dataset
**Ensure:** Validated dataset
1: repeat

### B. Dataset validation

As shown in Fig.8, raw generated dataset is not ideal to train sequence model, the error of the dataset will accumulate to the following steps. Especially for "pick" and "rotate left". The validation percentage is very low. To solve this problem and validate our dataset, we adopt KNN to validate incorrect data. First we find the center of each gesture dataset in the high dimensional space. Then assuming one sequence $Seq_{cand.}$ belongs to $Class_A$ and the distance to the center is $Seq_{cand.,A}$, we calculate it's all Euclidean distances $Seq_{cand.,class}$ to the center of other classes. If $Seq_{cand.,class} \leq Seq_{cand.,A}$, then label $Seq_{cand.}$ to usually vary for different environments. In this case, if a gesture is very fast completed and has $N_{less}(< M)$ frames, we assume no hand movements in residual data($M - N_{less}$). These residual data are very close to 0 and have no function on gesture recognition. Therefore, we can map $N_{less}$ frames to 30 to satisfy template matching using interpolating function. This alternative way is inevitable to bring some errors due to interpolating. If a gesture is performed slowly and has $N_{more}(> M)$ frames, we can only capture part of the gesture data due to that we have no prior knowledge to know when the gesture ends. So, we use incomplete data to perform gesture recognition.

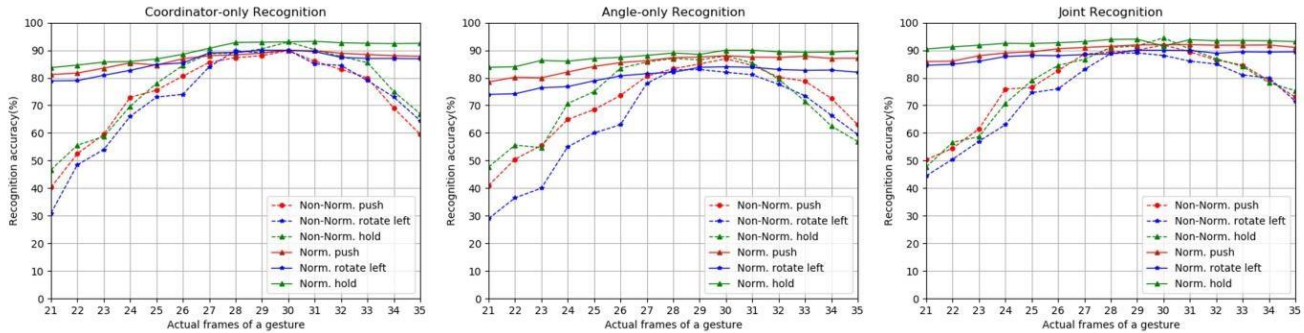

Fig. 9.     Recognition accuracy of three typical hand gestures under different actual frames, a)Coordinator-only approach, b)Angle-only approach, c)Joint

As shown in Fig.9, actual frames of a gesture denote the frame number when a gesture is completed. Here we take "push", "hold" and "rotate left" for example, each gesture is performed under coordinator-only mode, angle-only mode and joint mode, traversing 21-35 frames. We can see that the region with the highest precision is concentrated in the 30±1 frames, typically 30 frame owns the peak recognition accuracy. Besides, the influence of few frames on recognition accuracy is smaller than more frames.

Then we use normalization approach, the result is shown in Fig. 9. We can see that normalized gesture sequence has higher recognition accuracy than non-normalized. Besides, our proposed method performs much better on "More frames". There is no obvious decline in accuracy when actual frames are more than 30.

*2) Occluded frame number:* Occluded frame number is an important influence factor in gesture recognition, the ideal state is that all of the gesture frames can be captured. However, occlusion cannot be avoided in practical situations. As we settle in proposed dynamic matching method, least number required to perform gesture recognition should be more than 15 frames. Therefore, we observe the change of accuracy rate with 0 – 15 frames of data occlusion.

gesture while there is a cup (we use cup as occlusion object in this experiment) in the line-of-sight between the Kinect sensor and hands. Besides, we control the distance between the cup and the hands to ensure that there are more than 15 valid frames for each gesture. In this experiment, we don't record how many valid frames per movement, but just record the number of correctly identified. This is taken as average recognition accuracy.

As shown in Fig.10(a), the experiment is performed on "push" gesture based on joint recognition approach after normalization. We can find that with the increase of occlusion frames, the gesture recognition accuracy drops much faster. When it's over 15 frames, we stop recognizing gestures due to insufficient information. The other gestures owns a similar trend of accuracy change. Our method has better performance and robustness in low occlusion environment but also works in some high occlusion cases.

To validate the performance of our proposed method in occluded environment, we repeat 100 movements for each defined dynamic gestures. As for each captured keypoints sequence, we calculate its 3D Euclidean distances to 11

We compare our approach with DeepPrior [29] and M. M. et al. [11] in recognition accuracy and time efficiency. These two methods got promising results under occlusion. DeepPrior used Convolutional Neural Network approach and M. M. et al. used joint model-based and data-driven approach. But they are both designed for static gesture recognition. In order to compare the performance of dynamic gestures under the same conditions, we capture 30 frames of static data and form

*D. Comparison with related methods*

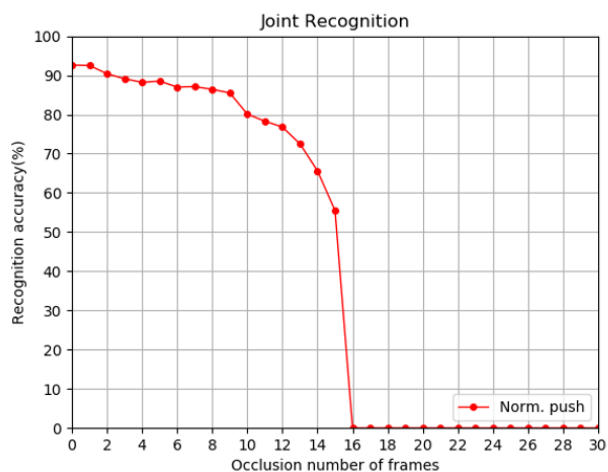

(a)

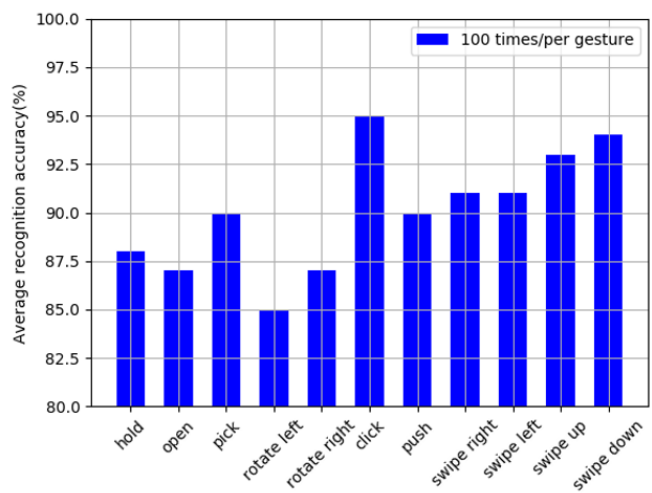

(b)

Fig. 10. (a) Recognition accuracy under different number frames of occlusion. (b) Average recognition accuracy of each gesture in 100 times.

As shown in Fig.10(b), we notice that except "hold","rotate left" and "rotate right", the other gestures can reach an high accuracy over 90%, and the four gestures also have an acceptable accuracy over 85%. This division may be caused by that each gesture depends on different keypoints and their corresponding number, some keypoints are easy to capture. For example, as for "click", it's highly dependent on the index finger, and the index finger is independent from the other keypoints and won't cause much errors. As for "rotate left/right", all keypoints are involved in gesture recognition and occupy a similar proportion. Some keypoints of data loss due to occlusion will bring more errors.

templates. Its category is determined by the minimum distance. we calculate and compare the average of the minimum distances of each gesture.

As illustrated in Fig.11(i), DeepPrior owns the largest average distances. This means it has the lowest accuracy. The reason could be that it does not have generalization on the diversity of gestures, especially on dynamic gestures. M. M. et al. is better than DeepPrior but still worse at sequence recognition. Also, since we supplement missing data and perform sequence normalization, actually we do not necessarily use 30 frames to identify a gesture. As shown in Fig.11. (ii), our approach is much better than DeepPrior and M. M. et al..

## VII. Conclusion

In this study, we propose a robust customized method for real-time hand tracking and gesture recognition with occlusion, which is very useful in practical interaction-based industrial applications. First, we analyze the hand kinematics and generate a key-point based representative model with relative coordinate vectors. Next, for deficient labeled data with occlusion, we introduce GAN to create a complete raw discrete sequence of hand movements. Furthermore, we build a time series prediction method in order to estimate the locations of key points on hands. At last, a sliding window matching method is applied to recognize hand gesture. In our experiment, our proposed method has higher recognition accuracy and less time cost than DeepPrior and M. Madadi et al.. Also, our method only needs less data to define and recognize a defined hand gesture. This solves the problem of insufficient data from practice.

It remains some improvements we can do in the future work. We will focus on more accurate extraction of key points, this will help improve the generated datasets greatly.

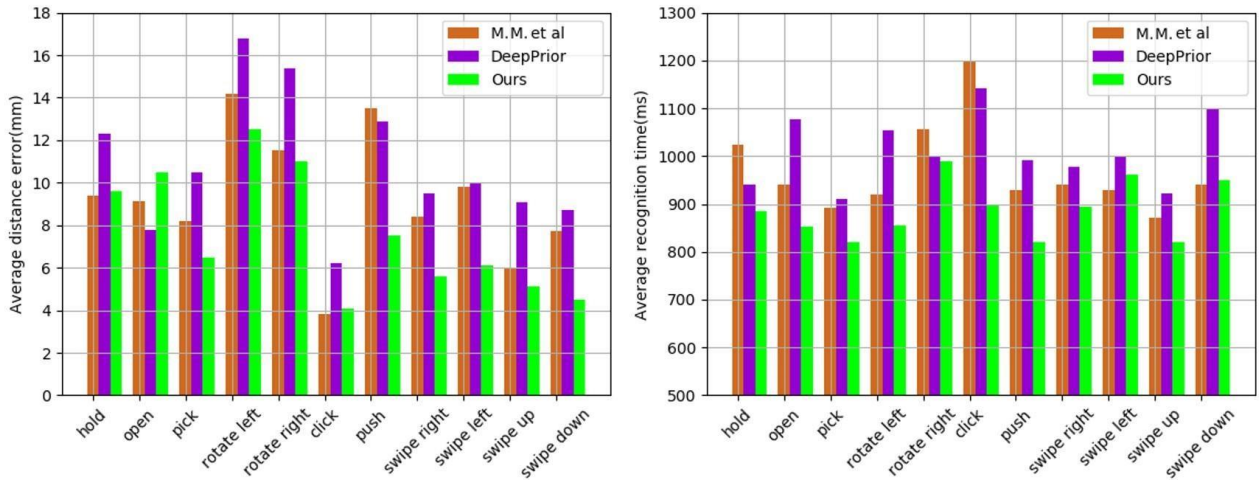

Fig. 11. Comparison of average distance error and recognition time of DeepPrior, M. M. et al. [11] and our method.

"Occlusion aware hand pose recovery from sequences of depth images," in *2017 12th IEEE International Conference on Automatic Face Gesture Recognition (FG 2017)*, May 2017, pp. 230–237.